# Electromagnetic Energy Balance Equations and Poynting Theorem

Gaobiao Xiao
Shanghai Jiaotong University, Shanghai, 200240 China
(e-mail: gaobiaoxiao@ sjtu.edu.cn).

**ABSTRACT** Poynting theorem plays a very important role in analyzing electromagnetic phenomena. The electromagnetic power flux density is usually expressed with the Poynting vector. However, since Poynting theorem basically focuses on the power balance in a system, it is not so convenient in some situations to use it for evaluating the electromagnetic energies. The energy balance issue for time varying fields is revisited in this paper, and a set of energy balance equations are introduced, and a modified method for evaluating power flux is proposed.

## I. INTRODUCTION

Poynting vector [1] is defined in terms of electric field and magnetic field and is widely accepted as the expression of power flux density. Poynting theorem is about the relationship between the Poynting vector and the electromagnetic energy densities, which provides an intuitive description of the electromagnetic energy propagation. However, the definition of electromagnetic flux density has been controversial [2]-[10], and some researchers have pointed out that there are limitations for Poynting theorem, though most of them have been ignored because of the great success of the wide application of Poynting theorem and Poynting vector [11]-[21].

It is known that Poynting theorem is not so efficient in handling issues concerning with reactive energies[8][9]. One famous example is the calculation of the reactive energies stored by antennas in an open space, which has been investigated for decades [22]-[31]. The difficulty comes from the fact that the total stored energy obtained by integrating the energy densities of $(0.5\vec{D}\cdot\vec{E})$ and $(0.5\vec{B}\cdot\vec{H})$ over the whole space is infinite, which is obviously unreasonable. Some researchers suggested that those fields associated with the propagating waves should not contribute to the stored reactive energies, and the results can become bounded by subtracting from the energy density an additional term associated with the radiation power [22][24]. However, it seems impossible to give a general definition for the term, because the propagation patterns are quite different for different antennas. It is proposed in [32] that the conventionally defined electric and magnetic energy densities, namely, $(0.5\vec{D}\cdot\vec{E})$ and $(0.5\vec{B}\cdot\vec{H})$, are generally for static fields and may not suitable for time-varying fields. Their integrations over a region are not rigorously equal to the total electromagnetic energy in that region in time varying situations. Since the energy densities involved in Poynting theorem are not absolutely correct for time varying fields, it is possible that limitations may exist for Poynting theorem, especially when reactive energies are concerned. No doubt that Poynting equation is rigorous because it is derived directly from Maxwell's equations. However, its interpretation can be slightly modified in some situations. It can be seen that Poynting theorem is basically directly describing the power balance in the system instead of the stored energy. Although it contains the energy densities, but it is their time varying rate that contributes to the balance. Therefore, it is not strange that sin some situations Poynting theorem is not much efficient for addressing issues concerning with total electromagnetic energies.

Based on these observations, I revisited the energy balance issue associated with current/charge sources in free space, and propose a set of balance equations for reactive electromagnetic energies. Because the energy balance equations are derived from the electromagnetic energies associated with given source distributions at a certain time, one cannot expect to get sufficient information for electromagnetic powers from these equations alone. Roughly speaking, the energy balance equations can be used complimentarily with Poynting theorem, i.e., the energy balance equations are used for handling energy balance problems, meanwhile, the Poynting equation is used for addressing power balance issues. Furthermore, a new formulation is proposed to calculate the power flow generated by current sources. Although no explicit expression for the modified power flux density is provided, it indeed gives a reasonable hypothesis that Poynting vector may not always exactly reflect the power flux density.

## II. GUIDELINES FOR MANUSCRIPT PREPARATION

Consider a time-varying charge distribution with density of $\rho(\vec{r},t)$ in region $V_a$ enclosed by surface $S_a$, as shown in Fig.1. A popular method to evaluate the total energy associated with the charge is to assume that all charges are moved piece by piece from infinitely far away to their current positions. Based on energy conservation law, it can be deduced that the total electric energy associated with the charge distribution is equal to the work done to them in the process of shifting them, which is derived to be [33][34]

$$W_\rho(t) = \frac{1}{2}\int_{V_a} \rho(\vec{r}',t)\phi(\vec{r}',t)d\vec{r}' \tag{1}$$

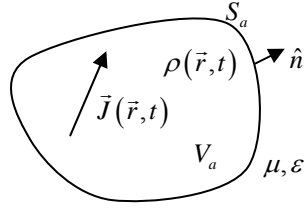

Fig. 1 Current and charge distribution enclosed by $S_a$.

The scalar potential $\phi(\vec{r},t)$ in three dimensional free space is expressed as

$$\phi(\vec{r},t) = \int_{V_a} \frac{\rho(\vec{r}',t-R/v)}{4\pi\varepsilon R} d\vec{r}' \qquad (2)$$

where $R = |\vec{r}-\vec{r}'|$ is the distance between the charge point and the evaluation point, $\varepsilon$ and $v$ represent the permittivity and propagation velocity in free space, respectively. Eq.(1) is well used for static fields. It is reasonable to postulate that it is still valid for time varying fields considered in classical electromagnetic problems, because (1) just describes the relationship between the electric energy and the charge distribution, as well as the scalar potential, at a certain time, no matter whether they are time-varying or not.

Similarly, if in region $V_a$ there is a current distribution with density of $\vec{J}(\vec{r},t)$, the magnetic energy associated with the current can be evaluated with

$$W_J(t) = \frac{1}{2}\int_{V_a} \vec{A}(\vec{r}',t)\cdot\vec{J}(\vec{r}',t) d\vec{r}' \qquad (3)$$

The vector magnetic potential $\vec{A}(\vec{r},t)$ in three dimensional free space is

$$\vec{A}(\vec{r},t) = \mu\int_{V_a} \frac{\vec{J}(\vec{r}',t-R/v)}{4\pi R} d\vec{r}' \qquad (4)$$

where $\mu$ is the permeability for free space. The zero point for both potentials is put at the infinity.

In time-varying situations, the electric field relates to both potentials with $\vec{E} = -\nabla\phi - \partial\vec{A}/\partial t$. Making use of vector identities, the electric energy in (1) can be transformed to

$$W_\rho(t) = \int_{V_a} \frac{1}{2}\left(\vec{E}\cdot\vec{D} + \frac{\partial\vec{A}}{\partial t}\cdot\vec{D}\right) d\vec{r}' + \frac{1}{2}\oint_{S_a} \phi\vec{D}\cdot\hat{r} dS \qquad (5)$$

where $\vec{D}(\vec{r},t)$ is the electric flux density, and $\hat{n}$ stands for the outward unit normal vector on the surface. Eq.(5) states that the total electric energy associated with the charge distribution is separated into two parts, one part is stored in the region $V_a$, expressed by the first term in the RHS of (5), and the rest part, expressed by the second term in the RHS of (5), will pass through the surface $S_a$ and leak to the region outside $V_a$. For the sake of convenience, a terminology of electric energy generation density is defined as

$$w_\rho(\vec{r},t) = \frac{1}{2}\rho(\vec{r},t)\phi(\vec{r},t) \qquad (6)$$

Meanwhile, define the stored reactive electric energy density as [32],

$$\tilde{w}_e(\vec{r},t) = \frac{1}{2}\left[\vec{E}(\vec{r},t)\cdot\vec{D}(\vec{r},t) + \frac{\partial\vec{A}(\vec{r},t)}{\partial t}\cdot\vec{D}(\vec{r},t)\right] \qquad (7)$$

and denote

$$\vec{\mathcal{E}}_e(\vec{r},t) = \frac{1}{2}\phi(\vec{r},t)\vec{D}(\vec{r},t) \qquad (8)$$

Here an upper script "~" is intentionally added on top of the parameter $\tilde{w}_e(\vec{r},t)$ to emphasize the difference compared to the definition for static fields. (5) is the balance equation for the electric energy associated with charge distribution $\rho(\vec{r},t)$. At a certain time $t$, one part of the electric energy is stored in region $V_a$ with energy density of $\tilde{w}_e(\vec{r},t)$, the other part is stored outside $V_a$, the amount of which can be calculated with the surface integral on $S_a$, however, the storing area and the energy density outside $V_a$ is not known. Vector $\vec{\mathcal{E}}_e(\vec{r},t)$ represents the energy passing through the surface per unit area. It is important to note that the balance equation is valid for any surface $S_a$ containing the charge distribution, including the spherical surface with its radius approaching infinity, hereafter denoted by $S_\infty$.

Following the same analogy, we can introduce the balance equation for the magnetic energy associated with current distribution $\vec{J}(\vec{r},t)$,

$$W_J(t) = \int_{V_a} \frac{1}{2}\left(\vec{B}\cdot\vec{H} - \vec{A}\cdot\frac{\partial \vec{D}}{\partial t}\right)d\vec{r}' + \frac{1}{2}\oint_{S_a}(\vec{H}\times\vec{A})\cdot\hat{n}dS \tag{9}$$

which implies that the magnetic energy is also divided into two parts. Define magnetic energy generation density as

$$w_J(\vec{r},t) = \frac{1}{2}\vec{A}(\vec{r},t)\cdot\vec{J}(\vec{r},t) \tag{10}$$

with the stored reactive magnetic energy density defined as

$$\tilde{w}_m(\vec{r},t) = \frac{1}{2}\left[\vec{B}(\vec{r},t)\cdot\vec{H}(\vec{r},t) - \vec{A}(\vec{r},t)\cdot\frac{\partial \vec{D}(\vec{r},t)}{\partial t}\right] \tag{11}$$

and denote

$$\vec{\mathcal{E}}_m(\vec{r},t) = \frac{1}{2}\vec{H}(\vec{r},t)\times\vec{A}(\vec{r},t) \tag{12}$$

(9) is the balance equation for the magnetic energy associated with current distribution at a certain time $t$. It can be interpreted in the same way as the electric energy balance equation (5).

It is worthwhile to emphasize that the densities of (7) (11) are defined only for reactive energies associated with sources. They are not defined to replace the ordinary energy densities for fields.

Combining (5) and (9), we get the balance equation for the total electromagnetic energy,

$$\int_{V_a}\left[w_\rho(\vec{r}',t)+w_J(\vec{r}',t)\right]d\vec{r}' = \int_{V_a}\left[\tilde{w}_e(\vec{r}',t)+\tilde{w}_m(\vec{r}',t)\right]d\vec{r}' + \oint_{S_a}\left[\vec{\mathcal{E}}_e(\vec{r}',t)+\vec{\mathcal{E}}_m(\vec{r}',t)\right]\cdot\hat{n}dS \tag{13}$$

Taking derivatives with respect to time yields

$$\frac{\partial}{\partial t}\int_{V_a}\left[w_\rho(\vec{r}',t)+w_J(\vec{r}',t)\right]d\vec{r}' = \frac{\partial}{\partial t}\int_{V_a}\left[\tilde{w}_e(\vec{r}',t)+\tilde{w}_m(\vec{r}',t)\right]d\vec{r}' + \frac{\partial}{\partial t}\oint_{S_a}\left[\vec{\mathcal{E}}_e(\vec{r}',t)+\vec{\mathcal{E}}_m(\vec{r}',t)\right]\cdot\hat{n}dS \tag{14}$$

which can be written in differential form

$$\frac{\partial}{\partial t}(w_\rho + w_J) = \frac{\partial}{\partial t}(\tilde{w}_e + \tilde{w}_e) + \nabla\cdot\vec{\mathcal{S}} \tag{15}$$

where

$$\vec{\mathcal{S}}(\vec{r},t) = \frac{\partial}{\partial t}(\vec{\mathcal{E}}_e + \vec{\mathcal{E}}_m) = \frac{\partial}{\partial t}\left(\frac{1}{2}\phi\vec{D} + \frac{1}{2}\vec{H}\times\vec{A}\right) \tag{16}$$

Eq.(14) describes the relationship between the varying rate of the total energy associated with the sources and the varying rate of the stored energies at time $t$. However, it is not sufficient to determine the total propagation power since it is basically not a power balance equation.

Reconsider the Poynting theorem, which directly comes from Maxwell's equations

$$\oint_{S_a}\vec{S}\cdot\hat{n}dS = -\int_{V_a}\vec{E}\cdot\vec{J}d\vec{r}' - \frac{\partial}{\partial t}\int_{V_a}\left(\frac{1}{2}\vec{D}\cdot\vec{E} + \frac{1}{2}\vec{B}\cdot\vec{H}\right)d\vec{r}' \tag{17}$$

where the Poynting vector is $\vec{S} = \vec{E}\times\vec{H}$. (17) is usually interpreted based on power conservation law. The RHS of (17) is the total power generated by $\vec{J}(\vec{r},t)$ in domain $V_a$, subtracting the increasing rate of energy stored in the domain, which should be the total radiated power coming out of the surface $S_a$. Hence, the Poynting vector is intuitively regarded as the power flux density. There is no doubt that Poynting equation itself is correct. However, the interpretation may be not so perfect due to the fact that the energy densities involved are basically only accurate for static fields. As illustrated by (5) and (9), for time varying fields, only the energy densities defined by (7) and (11) are strictly related to the stored energies associated with the current and charge sources. The time varying rate of the newly defined total reactive electromagnetic energy density can be rewritten as

$$\frac{\partial}{\partial t}(\tilde{w}_m + \tilde{w}_e) = \frac{\partial}{\partial t}\left(\frac{1}{2}\vec{B}\cdot\vec{H} + \frac{1}{2}\vec{D}\cdot\vec{E}\right) + \frac{1}{2}\left(\frac{\partial^2 \vec{A}}{\partial t^2}\cdot\vec{D} - \vec{A}\cdot\frac{\partial^2 \vec{D}}{\partial t^2}\right) \tag{18}$$

For time harmonic fields with time dependence of $\exp(j\omega t)$, we have $\partial^2/\partial t^2 \sim -\omega^2$, the second term in the RHS of (18) vanishes, and Poynting relationship can be written as

$$\oint_{S_a} \vec{S} \cdot \hat{n} dS = -\int_{V_a} \vec{E} \cdot \vec{J} d\vec{r}' - \frac{\partial}{\partial t}\int_{V_a} (\tilde{w}_m + \tilde{w}_e) d\vec{r}' \tag{19}$$

In this situation, the Poynting vector can be naturally considered as the power flux density since the energy densities are valid for both static and time varying cases. For non sinusoidal but slowly varying fields, the second term in the RHS of (19) can be very small because of small second order derivatives, therefore, (19) approximately holds true. In other situations, the Poynting vector may not correctly reflect the power flux density.

It is rational to introduce a modified formulation for the total power coming out from $S_a$ as

$$\begin{aligned}\tilde{P}_{rad}(t) &= -\int_{V_a} \vec{E} \cdot \vec{J} d\vec{r}' - \frac{\partial}{\partial t}\int_{V_a}(\tilde{w}_e + \tilde{w}_m) d\vec{r}' \\ &= -\int_{V_a} \vec{E} \cdot \vec{J} d\vec{r}' - \frac{\partial}{\partial t}\int_{V_a}\left(\frac{1}{2}\vec{B}\cdot\vec{H} + \frac{1}{2}\vec{D}\cdot\vec{E}\right) d\vec{r}' - \int_{V_a}\left(\frac{\partial^2 \vec{A}}{\partial t^2}\cdot\vec{D} - \vec{A}\cdot\frac{\partial^2 \vec{D}}{\partial t^2}\right) d\vec{r}' \\ &= \oint_{S_a} \vec{S}\cdot\hat{n} dS - \int_{V_a}\left(\frac{\partial^2 \vec{A}}{\partial t^2}\cdot\vec{D} - \vec{A}\cdot\frac{\partial^2 \vec{D}}{\partial t^2}\right) d\vec{r}' \\ &= -\int_{V_a}\vec{E}\cdot\vec{J} d\vec{r}' - \frac{\partial}{\partial t}\int_{V_a}\left[w_\rho(\vec{r}',t)+w_J(\vec{r}',t)\right] d\vec{r}' + \oint_{S_a}\vec{\mathcal{S}}\cdot\hat{n} dS' \\ &= -\int_{V_a}\left\{\vec{E}\cdot\vec{J} + \frac{\partial}{\partial t}\left(\frac{1}{2}\vec{A}\cdot\vec{J} + \frac{1}{2}\rho\phi\right)\right\} d\vec{r}' + \int_{S_a}\vec{\mathcal{S}}\cdot\hat{n} dS\end{aligned} \tag{20}$$

Note that in (20) the energy densities for time varying fields have been used to replace those for static fields. Eq. (20) shows that the total power contains two parts, the first part is the power generated by the source in the domain $V_a$, while the second part is the decreasing rate of the stored reactive energy in the domain. Substituting (17) and (18) into (20), we can show that

$$\tilde{P}_{rad}(t) = \oint_{S_a} \vec{S}\cdot\hat{n} dS - \int_{V_a}\frac{1}{2}\left(\frac{\partial^2 \vec{A}}{\partial t^2}\cdot\vec{D} - \vec{A}\cdot\frac{\partial^2 \vec{D}}{\partial t^2}\right) d\vec{r}' \tag{21}$$

As has been discussed right above, for time harmonic fields or slowly varying fields, the electromagnetic power coming out from a closed surface is just the total flux of Poynting vector over the surface.

Making use of (14) and (16), the total power can be expressed in terms of sources and the vector $\vec{\mathcal{S}}(\vec{r},t)$ as

$$\tilde{P}_{rad}(t) = -\int_{V_a}\left\{\vec{E}\cdot\vec{J} + \frac{\partial}{\partial t}\left(\frac{1}{2}\vec{A}\cdot\vec{J} + \frac{1}{2}\rho\phi\right)\right\} d\vec{r}' + \int_{S_a}\vec{\mathcal{S}}\cdot\hat{n} dS \tag{22}$$

### III. RADIATION PROBLEMS

For antennas in three dimensional free space, the stored reactive energies can be derived by letting $S_a \to S_\infty$, with the integration volume for reactive energies expanding to $V_\infty$. The source distribution region remains the same as $V_a$. Recalling that $\lim_{r\to\infty}(\vec{D}\cdot\hat{r}) \sim O(1/r^2)$, $\lim_{r\to\infty}\phi \sim O(1/r)$, where $\hat{r}$ is the unit radial vector, the surface integral at the RHS of (5) approaches zero, so the electric reactive energy is

$$W_e(t) = \int_{V_\infty}\frac{1}{2}\left(\vec{E}\cdot\vec{D} + \frac{\partial \vec{A}}{\partial t}\cdot\vec{D}\right) d\vec{r}' = \frac{1}{2}\int_{V_a}\rho\phi d\vec{r}' \tag{23}$$

Therefore, for radiation problems, the total reactive electric energy can be calculated in terms of fields with integration over the whole space, or in terms of charge and potential with integration over the source area, depending on which kind of information is available.

It is a little bit different for the magnetic energy. Since $\lim_{r\to\infty}(\vec{H}\times\vec{A})\cdot\hat{r} \sim O(1/r^2)$, the surface integral in (9) is usually a nonzero but bounded value. As has discussed in [32], this term can be considered as the energy stored at the infinity point beyond surface $S_\infty$, or equivalently, considered as being absorbed by the radiation resistor at infinity. The stored reactive magnetic energy accounts for the magnetic energy normally stored within $S_\infty$, and can be expressed with

$$W_m(t) = \int_{V_\infty}\tilde{w}_m d\vec{r}' = \int_{V_\infty}\frac{1}{2}\left(\vec{B}\cdot\vec{H} - \vec{A}\cdot\frac{\partial \vec{D}}{\partial t}\right) d\vec{r}' = \frac{1}{2}\int_{V_a}\vec{A}\cdot\vec{J} d\vec{r}' - \frac{1}{2}\oint_{S_\infty}(\vec{H}\times\vec{A})\cdot\hat{r} dS \tag{24}$$

Therefore, the basic way to calculate the reactive magnetic energy is to integrate the reactive magnetic energy density $\tilde{w}_m$ over the whole space. It can also be calculated with integration in terms of current and vector potential over the source area, subtracting the surface integral on $S_\infty$, i.e., the last term in (24). The radiation power is calculated using (22), with $S_a$ being replaced by $S_\infty$.

It is very important to notice that for pulse sources, the electromagnetic fields generated by the source will never reach $S_\infty$, so the surface integrals in (22) and (24) are zeros. Therefore, the stored reactive energies and radiation power in time domain for pulse excitations can be calculated with

$$\begin{cases} W_E(t) = \dfrac{1}{2}\int_{V_a} \rho(\vec{r}',t)\phi(\vec{r}',t)d\vec{r}' \\ W_m(t) = \dfrac{1}{2}\int_{V_a} \vec{A}(\vec{r}',t)\cdot\vec{J}(\vec{r}',t)d\vec{r}' \\ \tilde{P}_{rad}(t) = -\int_{V_a}\left\{\vec{E}\cdot\vec{J} + \dfrac{\partial}{\partial t}\left(\dfrac{1}{2}\vec{A}\cdot\vec{J} + \dfrac{1}{2}\rho\phi\right)\right\}d\vec{r}' \end{cases} \quad (25)$$

For time harmonic fields, it can be proved that $\oint_{S_\infty}\left(\vec{H}^*\times\vec{A}\right)\cdot d\vec{S} = 0$ [32]. Therefore, as has proved in [32], the time average energies and radiation power can be calculated with

$$\begin{cases} (W_E)_{av} = \mathrm{Re}\left\{\dfrac{1}{4}\int_{V_a}\rho(\vec{r}')\phi^*(\vec{r}')d\vec{r}'\right\} \\ (W_m)_{av} = \mathrm{Re}\left\{\dfrac{1}{4}\int_{V_a}\vec{A}(\vec{r}')\cdot\vec{J}^*(\vec{r}')d\vec{r}'\right\} \\ \tilde{P}_{rad} = -\mathrm{Re}\left\{\int_{V_a}\dfrac{1}{2}\vec{E}(\vec{r}')\cdot\vec{J}^*(\vec{r}')d\vec{r}'\right\} \end{cases} \quad (26)$$

where the same symbols are used for the corresponding phasors for the sake of convenience.

It can be seen from (25) and (26) that the reactive energies and the radiation power for pulses and harmonic sources can all be calculated with integrations over source region, which is very efficient for evaluating Q factors of large complex antennas.

## IV. CASE STUDY: HERTZIAN DIPOLE

A Hertzian dipole at the origin is analyzed to show the energy balance relationships. The moment of the dipole is assumed to be $ql\cos\omega t$, the scalar potential and the vector potential of which can be readily derived from a Hertzian potential $\Pi = (ql/4\pi r)\cos(\omega t - kr)$ [35]

$$\vec{A} = -\dfrac{\mu q l}{4\pi r}\sin(\omega t - kr)\left(\hat{r}\cos\theta - \hat{\theta}\sin\theta\right) \quad (27)$$

$$\varphi = \dfrac{ql}{4\pi\varepsilon}\cos\theta\left[\dfrac{1}{r^2}\cos(\omega t - kr) - \dfrac{k}{r}\sin(\omega t - kr)\right] \quad (28)$$

from which the fields are found to be

$$\vec{E} = \dfrac{k^2 ql}{4\pi\varepsilon r}\left\{\hat{r}2\cos\theta\left[\dfrac{1}{k^2 r^2}\cos(\omega t - kr) - \dfrac{1}{kr}\sin(\omega t - kr)\right] + \hat{\theta}\sin\theta\left[\left(\dfrac{1}{k^2 r^2} - 1\right)\cos(\omega t - kr) - \dfrac{1}{kr}\sin(\omega t - kr)\right]\right\} \quad (29)$$

$$\vec{H} = -\dfrac{\omega k q l}{4\pi r}\sin\theta\left[\dfrac{1}{kr}\sin(\omega t - kr) + \cos(\omega t - kr)\right]\hat{\varphi} \quad (30)$$

The total reactive energies at a certain time $t$ stored in the whole space outside a small sphere with radius $a$ can be derived from the fields and potentials as

$$W_m(t) = \alpha_0\left[1 - \cos 2(\omega t - ka)\right]\dfrac{1}{ka} + \alpha_0\left[\sin 2(\omega t - ka) - \lim_{r\to\infty}\sin 2(\omega t - kr)\right] \quad (31)$$

$$W_e(t) = \alpha_0\left[\dfrac{1}{k^3 a^3} + \dfrac{1}{ka} + \left(\dfrac{1}{k^3 a^3} - \dfrac{1}{ka}\right)\cos 2(\omega t - ka) - \dfrac{2}{k^2 a^2}\sin 2(\omega t - ka)\right] \quad (32)$$

and the nonzero surface integral is

9

$$\oint_{S_\infty} (\vec{\mathcal{E}}_e + \vec{\mathcal{E}}_m) \cdot \hat{r} dS = \alpha_0 \lim_{r\to\infty} \sin 2(\omega t - kr) \tag{33}$$

where $\alpha_0 = \frac{\omega^2 \mu k}{24\pi}(ql)^2$. Both the total reactive electric energy $W_m(t)$ and the total reactive magnetic energy $W_e(t)$ are bounded. As can be seen from (33) that although the surface integral is not zero, its time average is zero. Furthermore, we can show that

$$\oint_{S_\infty} \vec{\mathcal{S}} \cdot \hat{r} dS = 2\omega\alpha_0 \lim_{r\to\infty} \cos 2(\omega t - kr) \tag{34}$$

$$\oint_{S_\infty} \vec{S} \cdot \hat{r} dS = 2\omega\alpha_0 + 2\omega\alpha_0 \lim_{r\to\infty} \cos 2(\omega t - kr) \tag{35}$$

Comparing (34) with (35) implies that the DC component of the Poynting vector is cancelled in the energy balance equation. The time averages can be easily found and are listed below,

$$\begin{cases} P_{rad} = (S)_{av} = 2\omega\alpha_0 \\ (\tilde{W}_m)_{av} = \alpha_0 \left(\frac{1}{ka}\right) \\ (\tilde{W}_e)_{av} = \alpha_0 \left(\frac{1}{k^3 a^3} + \frac{1}{ka}\right) \end{cases} \tag{36}$$

The Q factor is calculated to be

$$Q = \frac{2\omega(\tilde{W}_e)_{av}}{P_{rad}} = \frac{1}{k^3 a^3} + \frac{1}{ka} \tag{37}$$

which is exactly in agreement with the results shown in [24].

For comparison, the equivalent circuit model proposed by Chu [34] for Hertzian dipole is show in Fig.2. Assuming that the current in the radiation resistor at the interface of $r = a$ is $i_R = I_0 \cos(\omega t - ka)$, the energies stored in the capacitor and the inductor can be derived to be

$$\begin{cases} W_C = \frac{I_0^2}{2\omega}\left[\frac{1}{ka} + \frac{1}{(ka)^3} + \left(\frac{1}{(ka)^3} - \frac{1}{ka}\right)\cos 2(\omega t - ka) - \frac{1}{(ka)^2}\sin 2(\omega t - ka)\right] \\ W_L = \frac{I_0^2}{2\omega}\left(\frac{1}{ka}\right)\sin^2(\omega t - ka) \end{cases} \tag{38}$$

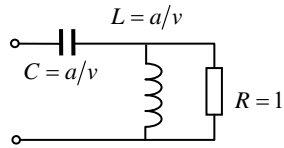

Fig.2 Equivalent circuit model for Hertzian dipole radiation.

If we let $I_0^2 = 2\omega\alpha_0$, it can be checked that $W_C(t) = W_e(t)$, and $W_L(t) \approx W_m(t)$. It is because that in the equivalent circuit, the radiation resistor is assumed to be connected at the input port, namely, at $r = a$. However, for real antennas, the radiation resistor is at $r \to \infty$. For time-averaged energies, the equations hold strictly.

**V. CONCLUSION**

With the energy densities for time varying fields proposed in [32], the balance equations for reactive energies associated with charge or current sources are created for fields at a certain time. These balance equations are for the stored energies instead of power flux. The formulation proposed for calculating the electromagnetic power is slightly different from the conventional method based on Poynting theorem, the detailed derivation shows that the modification may be a possible way for remedying the limitations of applying Poynting vector for power flux density.